# The future of generative AI chatbots in higher education

Joshua Ebere Chukwuere

Department of Information Systems, North-West University, South Africa

joshchukwuere@gmail.com

**Abstract**

The integration of generative Artificial Intelligence (AI) chatbots in higher education institutions (HEIs) is reshaping the educational landscape, offering opportunities for enhanced student support, and administrative and research efficiency. This study explores the future implications of generative AI chatbots in HEIs, aiming to understand their potential impact on teaching and learning, and research processes. Utilizing a narrative literature review (NLR) methodology, this study synthesizes existing research on generative AI chatbots in higher education from diverse sources, including academic databases and scholarly publications. The findings highlight the transformative potential of generative AI chatbots in streamlining administrative tasks, enhancing student learning experiences, and supporting research activities. However, challenges such as academic integrity concerns, user input understanding, and resource allocation pose significant obstacles to the effective integration of generative AI chatbots in HEIs. This study underscores the importance of proactive measures to address ethical considerations, provide comprehensive training for stakeholders, and establish clear guidelines for the responsible use of generative AI chatbots in higher education. By navigating these challenges, and leveraging the benefits of generative AI technologies, HEIs can harness the full potential of generative AI chatbots to create a more efficient, effective, inclusive, and innovative educational environment.

**Keywords:** Academics, AI chatbots, Artificial Intelligence (AI), ChatGPT, Generative AI chatbots, Higher education institutions (HEIs), Higher education

## Introduction

The application of Artificial Intelligence (AI) technologies, particularly generative AI chatbots, into higher education institutions (HEIs) is reshaping the landscape of academic environments worldwide. As HEIs navigate the complexities of a rapidly evolving technological era, the future of AI chatbots in higher education presents both unprecedented opportunities and challenges (Sain & Hebebci, 2023). With the emergence of generative AI chatbots like ChatGPT, HEIs are at the forefront of a technological revolution that promises to revolutionize traditional teaching and learning, and research paradigms. Generative AI has the power to revolutionize how things are done and it is classified into text-to-text, text-to-audio, text-to-video, and many more (Aydin & Karaarslan, 2023). A generative AI chatbot is a conversational AI tool built on natural language processing (NLP) and deep learning techniques with the ability to mimic human writing style, and abilities, and create human-like text in real-time. It generates text that resembles that of a human. For example, ChatGPT is regarded as a generative AI chatbot with the ability to generate eloquent and logical human-like text and conservations in real time (Kim, Kim & Baek, 2024). The utilization of generative AI chatbots in educational settings opens new avenues for interactive learning, innovative assessment methods, and data-driven insights for continuous improvement within the academic ecosystem. However, alongside these promising advancements, HEIs also face a myriad of challenges in harnessing the full potential of AI chatbots while safeguarding academic integrity, addressing ethical concerns, and ensuring equitable access to educational resources (Mahmud, 2024).

As HEIs strive to adapt to the increasing technological dependence on AI tools, it becomes imperative to explore the multifaceted implications of AI chatbots in higher education (Dempere, Modugu, Hesham & Ramasamy, 2023). This study aims to delve into the prospects of generative AI chatbots in HEIs, examining the opportunities they present for enhancing student learning outcomes, academic productivity, institutional efficiency, and research processes. Furthermore, the study seeks to address the critical challenges that HEIs may encounter in the application of generative AI chatbots, such as academic integrity concerns, user input understanding, ethical considerations, and privacy implications.

By conducting a comprehensive analysis of the future trajectory of generative AI chatbots in higher education, this study endeavors to provide valuable insights into the evolving educational landscape and propose strategic solutions to mitigate potential risks and maximize the benefits of generative AI chatbots. Through a nuanced exploration of the opportunities, challenges, and implications of generative AI chatbots in HEIs, this research aims to contribute to the ongoing discourse on the integration of AI, especially generative AI chatbots in higher education and pave the way for informed decision-making and effective implementation strategies in the digital age of academia.

**Problem statement**

The future higher education institutions (HEIs) are turning towards AI technological-driven. The future where technology is the center of educational settings and operations. According to Ahmad (2020) and Mustar and Wright (2010), HEIs will be technologically dependent in the future. The technological dependence of higher education will be driven by AI technology. For example, the emergence of ChatGPT in November 2022 has redefined how student assessments and research are conducted and raised issues of ethical, privacy, and academic integrity concerns. However, the future of generative AI chatbots like ChatGPT is uncertain in higher education (Elbanna & Armstrong, 2024). The uncertainty means more research into the potential future of AI and understanding the current dual impact in higher education institutions (HEIs).

Generative AI chatbot applications in higher education present positive and negative consequences. According to Ilieva, Yankova, Klisarova-Belcheva, Dimitrov, Bratkov and Angelov (2023), AI chatbots can promote critical thinking skills through interactive conversations that prompt students to analyze information and consider diverse viewpoints. The author further suggests that AI tools such as ChatGPT provide support in various academic tasks, such as idea generation, research, and writing, streamlining academic processes and enhancing student and academic productivity. However, academicians advise that HEIs to develop a flexible assessment evaluation strategy to accommodate the AI chatbots (Adeshola & Adepoju, 2023).

Furthermore, the risk of bias and dissemination of false information by AI chatbots raises concerns about the accuracy and reliability of educational content (Ilieva et al., 2023). Inadequate assessment design involving generative AI chatbots may also impact the validity and effectiveness of student evaluations, requiring careful consideration in educational settings. Despite the potential of AI chatbots in higher education, the dual effect on student experience and performance is conflicting (Wu & Yu, 2024). Then, this study aimed to examine existing literature on the future of generative AI chatbots in higher education institutions (HEIs). The findings from the study promote a better understanding of the future, opportunities, challenges, and strategies to mitigate the challenges of generative AI chatbots in higher education.

**Research objectives**

The application of generative AI chatbots continues to be an evolving discussion and a major concern in higher education. This study is guided by these objectives:

⟩ To explore the future of generative AI chatbots in higher education,

⟩ To explore the challenges higher education may face in the future in applying generative AI chatbots,

⟩ To propose solutions to mitigating the identified challenges.

**Research methodology - Narrative Literature Review (NLR)**

Several literature reviews can be used to understand the future of generative AI chatbots in higher education. Among them is the narrative literature review (NLR). According to Chukwuere (2023), the NLR methodology is discussed as a valuable approach for conducting exploratory research. It explores publications that explain the current research topic and existing viewpoints (Rother, 2007). NLR is highlighted for its flexibility in synthesizing existing research on a specific topic, identifying gaps in knowledge, and providing a comprehensive understanding of the subject matter (Chukwuere, 2023; Byrne, 2016). However, it is noted that narrative reviews may lack rigor and replicability (Chukwuere, 2023), and there is a risk of bias and subjectivity in the analysis process. It implies that NLR has no specified steps or guidelines.

A narrative literature review (NLR) is used in exploratory research as applied as demonstrated in the study by Jimenez, Gray, Di Michele, Said, Reed and Kench (2023) in the exploration of educational interventions in Medical Radiation Sciences. The methodology allows for a broad examination of the current evidence, providing a holistic view of the topic. This methodology enables the researchers to synthesize information from diverse sources and present a cohesive narrative that highlights key themes and findings. Additionally, a narrative review can help identify gaps in the existing literature, suggest areas for further research, and offer insights for practice based on the collective knowledge available. The collection of knowledge involves a search in the synthesizing of findings from peer-reviewed materials (Green, Johnson & Adams, 2006). By incorporating a narrative approach, researchers can effectively communicate complex information, facilitate understanding, and contribute to the ongoing development of educational practices in the field of medical radiation sciences (Jimenez et al., 2023).

In this study, NLR assisted the author in examining the future of generative AI chatbots in higher education by sourcing existing academic material across different databases like Google Scholar, Scopus, and many others using pre-defined keywords. The author identified keywords that guided the search, for example, the keyword includes but are not limited to "academic use of AI chatbots", "artificial Intelligence (AI) in higher education", "ChatGPT", "generative AI chatbots use in higher education", "higher education institutions (HEIs)", and "higher education". The sourced materials enable the author to identify research gaps, current discussions, and challenges in the topic and able the researcher to propose mitigating solutions and strategies to address the challenges.

**Eligibility and criteria**

The ability to search for the right academic articles depends on the set eligibility and criteria. According to Dehkordi, Mazaheri, Ibrahim, Dalvand and Gheshlagh (2021), at the time the research questions and potential database to source the paper are identified, the searching methods, screening criteria, eligibility checks, data sourcing, and analysis will be specified. The eligibility of any article can be subjected to some period like in months and years, in this case, articles published between 2021 to February 2024 were considered. Also, when the initial article screening ends, potential articles will undergo a selection process through inclusion and exclusion criteria (Kumarasamy, Sabarimurugan, Madurantakam, Lakhotiya, Samiappan, Baxi & Jayaraj, 2019). The inclusion and exclusion criteria assist the researcher in avoiding bias in choosing a given article (Table 1).

**Table 1:** Inclusion and exclusion

|   | Inclusion | Exclusion |
|---|---|---|
| 1 | Empirical research papers (articles) | Abstract missing |
| 2 | The paper is written in the English language | Paper not written in the English language |
| 3 | Articles that cover that keyword | Papers with citations and references |
| 4 | Published articles between 2021 and February 2024 | Articles that don't focus on the keywords. |

**Addressing the research objectives**

*The future of generative AI chatbots in higher education*

The future of generative AI chatbots in higher education is a topic of growing interest and concern due to the potential benefits and challenges they present. The future of generative AI chatbots in higher education is poised to revolutionize the academic landscape by offering innovative solutions to enhance student learning experiences, streamline administrative processes, and improve research writing. The future potential and application are strengthened by its effects on students' learning experience, outcomes, and performance (Wu Yu, 2024). Generative AI chatbots, powered by artificial intelligence (AI) algorithms, have the potential to provide personalized support to students, offering instant responses to queries related to course information, schedules, assignments, and academic resources (Rathore, 2023). As highlighted in the article by Grassini (2023), the integration of AI applications like ChatGPT in educational settings can significantly enhance the learning experience for students and increase academic research engagement and publication 24/7. With their 24/7 accessibility, generative AI chatbots can ensure that students and academics have access to assistance at any time, catering to the needs of distance learning and those with varying schedules and learning needs and expectations. Moreover, these AI-driven assistants can automate routine administrative tasks such as admissions inquiries, course registrations, fee payments, and appointment scheduling, freeing up staff members to focus on more strategic initiatives (Rathore, 2023).

In addition to improving student learning, academic support, research writing, and administrative efficiency, generative AI chatbots can further facilitate personalized learning experiences by recommending relevant study materials, providing feedback on assignments, and offering tailored learning pathways based on individual student needs and preferences. By collecting and analyzing data on student interactions and performance, AI chatbots enable HEIs to gain valuable insights into student behavior and engagement levels, allowing for targeted interventions and enhancements in the learning process (Kumar, Rao, Singhania, Verma & Kheterpal, 2024; Rathore, 2023; George & Wooden, 2023). Furthermore, the integration of generative AI chatbots with learning management systems (LMS) can provide seamless access to course content, assessments, and collaborative tools, enriching the overall learning experience for both students and lecturers.

As generative AI chatbots continue to evolve and become more sophisticated, they hold the potential to support higher education institutions (HEIs) and academics by providing information on institutional policies, faculty resources, professional development opportunities, and administrative procedures. By streamlining communication within academic departments and across different campus units, generative AI chatbots can foster a more connected and efficient academic environment. The future of generative AI chatbots in higher education is characterized by their ability to enhance student support, personalized learning experiences, streamline administrative tasks, support research writing, and provide valuable insights for continuous improvement in the academic ecosystem.

*The challenges higher education may face in the future in applying generative AI chatbots*

In the realm of higher education, the application of generative AI chatbots presents several challenges that HEIs may face in the future. Addressing these challenges below will be essential for HEIs to harness the full potential of generative AI chatbots while mitigating risks and ensuring a positive impact on teaching, student learning outcomes/performance, and academic research publications. Also, these challenges are crucial to address to ensure the effective integration of AI technologies in educational settings (Grassini, 2023; Følstad, Araujo, Law, Brandtzaeg, Papadopoulos, Reis & Luger, 2021 Grassini, 2023):

**Academic integrity concerns:** One of the primary challenges is maintaining academic integrity in an environment where generative AI chatbots can potentially be used for plagiarism or academic dishonesty. Higher education institutions (HEIs) together with academics must develop strategies to detect and prevent misuse of AI-generated content in student and academic work to uphold academic standards. However, this academic integrity concern remains a challenging aspect of the application of generative AI chatbots in this modern age of education.

**User input understanding:** Understanding user input remains a challenge for generative AI chatbots. Even though natural language understanding and purpose prediction have improved with machine learning techniques, conversational failures in generative AI chatbot conversations can occur because of misunderstandings, particularly in daily or real-world contexts.

**Adaptation to technological changes:** Higher education institutions must adapt to the rapid pace of technological advancements in AI. Keeping up with evolving AI technologies and trends requires continuous learning and flexibility to incorporate new tools and methodologies into existing educational practices.

**Detection of AI-generated content:** As AI technology advances, detecting AI-generated content becomes increasingly difficult. HEIs and even academics may struggle to differentiate between human-written, and AI-generated text, necessitating the development of more sophisticated detection tools to ensure the authenticity of student and academic research work.

**Training and skill development:** Academics require training to effectively utilize generative AI chatbots in educational settings. Developing the skills to leverage generative AI tools for enhancing teaching practices, student learning and research outcomes is essential but may require significant investment in professional development programs.

**Ethical and privacy concerns:** The use of generative AI chatbots raises ethical considerations related to data privacy, algorithmic bias, and the responsible use of technology in higher education. HEIs must establish clear guidelines and policies to address these ethical and privacy concerns, safeguard student data, and uphold academic integrity.

**Adapting to user and conversational context:** Generative AI chatbots need to adapt to the user and conversational context effectively. This is crucial, especially in sensitive scenarios such as in the health domain, where generative AI chatbots must adjust the conversation to the social, emotional, and health literacy aspects of users. Challenges persist in modeling and adapting to the user and conversational context.

**Inclusive and responsible design:** Generative AI chatbots face challenges related to bias and inclusion. Systematic research on the inclusive and universal design of generative AI chatbots is lacking. Understanding the many language components of discourse as well as being aware of larger social and cultural issues are necessary for inclusive and ethical design. To guarantee universal design and lessen prejudice, research is required on the usage of generative AI chatbots and obstacles to onboarding.

**Resource allocation:** Implementing generative AI chatbots in higher education requires substantial resources in terms of infrastructure, training, and ongoing support. HEIs may face challenges in allocating sufficient resources to ensure investment, the successful integration, and maintenance of generative AI technologies.

**Adaptation to technological changes:** Higher education institutions must adapt to the rapid pace of technological advancements in generative AI chatbots. Keeping up with evolving AI technologies and trends requires continuous learning and flexibility to incorporate new tools and methodologies into existing educational practices.

*Emerging challenges and benefits of generative AI Chatbots in higher education*

The integration of generative AI-based chatbots like ChatGPT into higher education presents a range of emerging challenges and benefits. One significant challenge identified is the potential ambiguity surrounding the rules and guidelines for using ChatGPT in academic settings. Students and even academics may struggle with understanding the boundaries of acceptable use, leading to uncertainties in how to incorporate the tool effectively into their learning process (Neumann, Rauschenberger & Schön, 2023). Issues such as academic misconduct, including plagiarism and cheating, pose ethical challenges that HEI and academics must address to maintain academic integrity (Ilieva et al., 2023). Additionally, the heterogeneous evaluation practices among lecturers (academics) can further complicate matters, as inconsistencies in assessing the use of ChatGPT may create confusion for students regarding what is permissible in their academic work (Neumann et al., 2023).

Another critical challenge is determining the acceptable and unacceptable use of generative AI chatbots like ChatGPT in educational contexts. This issue involves various considerations, including the risk of plagiarism and cheating, particularly in assessments and research where traditional plagiarism detection tools may not effectively identify text generated by ChatGPT (Richards, Waugh, Slaymaker, Petre, Woodthorpe & Gooch, 2024; Neumann et al., 2023; Jarrah, Wardat & Fidalgo, 2023). Moreover, the integration of generative AI chatbots like ChatGPT into teaching and assessment methods can pose a time-consuming task for academics, requiring adjustments to course materials and evaluation processes to accommodate the use of the tool effectively (Neumann et al., 2023; Liu, Ren, Nyagoga, Stonier, Wu & Yu, 2023).

Despite these challenges, the adoption of generative AI chatbots like ChatGPT in higher education also offers several benefits. One notable advantage is the potential enhancement of the virtual tutoring system through personalized support for students. By leveraging ChatGPT as a virtual tutor, students can receive tailored explanations, translations, or verification of artifacts, thereby improving their fundamental understanding of specific topics (Sohail, Farhat, Himeur, Nadeem, Madsen, Singh & Mansoor, 2023; Neumann et al., 2023). Furthermore, the innovation potential of generative AI tools like ChatGPT presents opportunities for introducing novel teaching approaches and strategies. Integrating ChatGPT into pedagogical methods such as problem-based learning or flipped classrooms can diversify learning experiences for students and foster creativity in educational practices (Neumann et al., 2023).

*The proposed solutions for mitigating the challenges*

The increasing sophistication of AI technology raises concerns about the potential misuse of AI-generated content, particularly in academic settings. There is a growing need for academics to be equipped with the skills to detect AI-generated text in student work (Fleckenstein, Meyer, Jansen, Keller, Köller & Möller, 2024; Otterbacher, 2023; Grassini, 2023). Additionally, academics must be educated on how to maximize the potential of AI tools like ChatGPT in lesson preparation and evaluation (Grassini, 2023).

Furthermore, Labadze, Grigolia and Machaidze (2023), and Grassini (2023) emphasize the importance of establishing clear guidelines for the acceptable use of AI chatbots in higher education. Academics are encouraged to incorporate digital-free components into evaluation tasks to ensure that students demonstrate their competencies without relying on external tools (Grassini, 2023). This approach aims to maintain academic integrity and ensure fair assessment practices in the face of advancing AI technology. Overcoming current and future challenges confronting the implementation of generative AI chatbots in higher education involves some mitigating solutions. Table 2 presents several proposed solutions and strategies for mitigating the challenges of using generative AI chatbots in higher education now and in the future.

**Table 2:** Solutions and strategies for mitigating the challenges in the application of generative AI chatbots

|   | Solutions for mitigations | Descriptions |
|---|---|---|
| 1 | Developing ethical principles | Setting ethical codes and principles for using generative AI chatbot systems in education is crucial to address ethical concerns and ensure responsible AI deployment. Researchers and stakeholders should work towards defining ethical guidelines that govern the use of generative AI chatbots in educational settings (Okonkwo & Ade-Ibijola, 2021). |
| 2 | Education and training | It is essential to provide comprehensive education and training to HEI management, academics, and students. This involves educating people on the advantages, restrictions, and possible dangers of generative AI chatbots through workshops, |

| | | seminars, webinars, or online courses. Data ethics, digital literacy, critical thinking, plagiarism, responsible AI usage (for instance, ChatGPT), and academic integrity should all be included (Michel-Villarreal, Vilalta-Perdomo, Salinas-Navarro, Thierry-Aguilera & Gerardou, 2023). |
|---|---|---|
| 3 | Collaboration, and interdisciplinary efforts | Emerging threats necessitate a multidisciplinary response. In order to solve the issues, HEIs can promote cooperation between the departments of computer science, ethics, education, and law, among other subjects (Michel-Villarreal et al., 2023). Institutions may successfully identify and address the problems connected with the incorporation of generative AI chatbots in higher education by promoting multidisciplinary debates and partnerships. |
| 4 | Improving usability testing | Enhancing the usability and trustworthiness of generative AI-based chatbot systems through rigorous testing and evaluation is essential. Assessing the effects of generative AI chatbot use on students' learning abilities and experiences can help improve the design and functionality of generative AI chatbots in education (Okonkwo & Ade-Ibijola, 2021). |
| 5 | Research and development | It is essential to fund research and development projects centred on generative AI chatbots in higher education. According to Michel-Villarreal et al. (2023), this entails researching the effects of generative AI chatbots on student learning outcomes and experiences, creating algorithms for bias detection and mitigation, looking into ways to integrate generative AI chatbots with human expertise, and looking into ethical frameworks for the use of generative AI chatbots in educational settings. Best practices and policy choices can be guided by evidence-based insights from research. |
| 6 | Enhancing technical advancements | Focusing on technical advancements, such as expanding knowledge banks, automating tests, and advancing generative AI chatbot features, can improve the functionality and intelligence of generative AI chatbots in educational contexts (Okonkwo & Ade-Ibijola, 2021). More research and development efforts should be directed towards enhancing the technical capabilities of generative AI chatbots for educational purposes. |
| 7 | Human oversight and intervention | Generative AI chatbots can be very helpful, but human supervision and involvement are still necessary. According to Michel-Villarreal et al. (2023), higher education institutions should place a strong emphasis on the guidance, correctness verification, and provision of extra context that academics and instructors provide to students. Human interaction can assist in addressing constraints, confirming facts, and guaranteeing the calibre of instruction. |
| 8 | Empirical studies | There is a call for more empirical studies focusing on AI technologies (for example, ChatGPT) in real teaching and learning settings to serve educational needs and purposes (Zhang & Aslan, 2021). Conducting rigorous research in actual educational environments can provide valuable insights into the effectiveness of generative AI chatbots applications and help tailor solutions to meet specific educational goals. |

| 9 | Educational perspectives | The dearth of educational viewpoints in generative AI chatbot research and development has been brought to light by recent publications (Zhang & Aslan, 2021). It is crucial to include educators and educational researchers in the process of technology innovation in order to solve this difficulty. By integrating input from higher education settings, AI technologies (generative AI chatbots) can be designed to better meet the needs and preferences of students and educators. |
| --- | --- | --- |
| 10 | Informed consent and opt-out options | According to Michel-Villarreal et al. (2023), academics and students ought to be offered the option to interact with generative AI chatbots (models) or decline if they have reservations or would like to use other methods of instruction and research. HEIs should make sure that students are fully informed about the goals and consequences of employing AI technologies so that they may choose whether or not to participate. |
| 11 | Meaningful communication | Meaningful communication among stakeholders with different areas of expertise, perspectives, and backgrounds is crucial for promoting generative AI chatbots (Zhang & Aslan, 2021). By fostering collaboration and dialogue between technology experts, educators, administrators, and researchers, institutions can develop actionable guidelines and effective strategies for implementing generative AI chatbots in higher education. |
| 12 | Continuous monitoring and evaluation | HEIs should keep an eye on and assess generative AI chatbot usage on a regular basis to spot any possible moral or instructional concerns. According to Michel-Villarreal et al. (2023), this entails gathering input from teachers and students, evaluating the influence on learning objectives, and making the required modifications to guarantee appropriate usage and efficacy. |

By adopting these strategies in Table 1, HEIs can promote the ethical and responsible use of generative AI chatbots in higher education, ensuring ethical standards are upheld, student and academic rights are protected, and the overall learning experience and research writing are enhanced.

**Contributions of the study**

The study on the future implications of generative AI chatbots in higher education institutions makes several novel contributions to the existing literature. Firstly, it offers a comprehensive exploration of the opportunities and challenges presented by generative AI chatbots in HEIs. By delving into the potential impact of these technologies on teaching, learning, and research processes, the study provides a nuanced understanding of how generative AI chatbots can reshape the educational landscape. Secondly, the study contributes strategically by proposing solutions to maximize the effectiveness of generative AI chatbots while addressing ethical and privacy concerns. This proactive approach to navigating the ethical implications of AI technologies in higher education sets the groundwork for responsible AI integration and underscores the importance of establishing clear guidelines for their use.

Moreover, the study synthesizes existing research on generative AI chatbots in higher education through a narrative literature review (NLR) methodology. By consolidating diverse perspectives and findings from academic databases and scholarly publications, the study offers a holistic view of the subject matter, highlighting key insights and trends in the field. Additionally, the identification of key challenges such as academic integrity concerns, user input understanding, and resource allocation represents a significant contribution to the study. By pinpointing these obstacles, the research sheds light on the complexities that HEIs may encounter in effectively integrating generative AI chatbots into educational settings, prompting further discussion and exploration in these areas.

Lastly, the study emphasizes the importance of responsible generative AI use in higher education by advocating for ethical guidelines, stakeholder training, and a proactive approach to addressing ethical and practical considerations. By promoting responsible generative AI integration, the study aims to guide HEIs toward leveraging the benefits of generative AI chatbots while mitigating risks and ensuring a positive impact on teaching, learning outcomes, and academic research publications.

**Conclusions**

The study on the future implications of generative AI chatbots in higher education institutions sheds light on the transformative potential of AI technologies in reshaping teaching, learning, and research processes within HEIs. By identifying key opportunities and challenges associated with generative AI chatbots, the research underscores the importance of responsible AI integration and proactive measures to address ethical concerns in educational settings. Looking ahead, future studies in this field could delve deeper into specific aspects of generative AI chatbots in higher education, such as the impact on student learning outcomes, the effectiveness of AI-generated content in academic settings, and the role of human oversight in AI-supported education. Additionally, research focusing on the development of advanced AI algorithms for bias detection, the integration of generative AI chatbots with human expertise, and the exploration of ethical frameworks for generative AI chatbot use in higher education can provide valuable insights for policy decisions, formulation, and best practices.

Furthermore, future studies could explore the long-term implications of generative AI chatbots on the academic landscape, including their influence on pedagogical approaches, assessment practices, and the overall student experience. By conducting empirical studies in real teaching and learning environments, researchers can gather evidence-based insights to inform the responsible integration of AI technologies in higher education. This study sets the stage for continued exploration and innovation in the field of generative AI chatbots in higher education, offering a foundation for future research endeavors that aim to enhance teaching and learning practices, promote ethical generative AI use, and ensure the positive impact of technology on the academic community.